\documentclass[  
reprint, 
aps,
amsmath,
amssymb,
amsfonts, 
superscriptaddress,
prx,
longbibliography
]{revtex4-2} 

\usepackage{braket}
\usepackage{graphicx}
\usepackage{bm}
\usepackage[colorlinks=true,allcolors=blue]{hyperref} 
\usepackage[T1]{fontenc}
\usepackage{amsmath}

\newcommand{\er}{Er$^{3+}$}
\newcommand{\ercawo}{Er$^{3+}$:CaWO$_4$}

\date{\today}

\begin{document}
\title{Rephasing spectral diffusion in time-bin spin-spin entanglement protocols}

\author{Mehmet T.~Uysal}
\affiliation{\mbox{Department of Electrical and Computer Engineering, Princeton University, Princeton, NJ 08544, USA}}

\author{Jeff D.~Thompson}
\email[]{jdthompson@princeton.edu}
\affiliation{\mbox{Department of Electrical and Computer Engineering, Princeton University, Princeton, NJ 08544, USA}}

\begin{abstract}
Generating high fidelity spin-spin entanglement is an essential task of quantum repeater networks for the distribution of quantum information across long distances. Solid-state based spin-photon interfaces are promising candidates to realize nodes of a quantum network, but are often limited by spectral diffusion of the optical transition, which results in phase errors on the entangled states. Here, we introduce a method to correct phase errors from quasi-static frequency fluctuations after the entangled state is generated, by shelving the emitters in the excited state to refocus the unknown phase.
For quasi-static frequency fluctuations, the fidelity is determined only by the lifetime of the excited state used for shelving, making it particularly suitable for systems with a long-lived shelving state with correlated spectral diffusion. Such a shelving state may be found in Kramers doublet systems such as rare-earth emitters and color centers in Si or SiC interfaced with nanophotonic cavities with a strongly frequency-dependent Purcell enhancement. The protocol can be used to generate high-fidelity entangled spin pairs without reducing the rate of entanglement generation.
\end{abstract}

\maketitle


Quantum networks are an enabling technology for secure communication \cite{gisin2002quantum}, distributed quantum computing and quantum metrology \cite{Awschalom2021}. Entanglement between distant spin-photon interfaces, as the basic link of a long distance network, has been demonstrated with trapped ions \cite{blinov2004observation}, neutral atoms \cite{volz2006observation,wilk2007single}, quantum dots \cite{gao2012observation,de2012quantum}, and atom-like defects in the solid-state \cite{togan2010quantum,bernien2013heralded,Bhaskar2020}.
Solid-state systems are promising for realizing large-scale networks with on-chip integration, but need to contend with spectral diffusion, frequency fluctuations of the optical transitions owing to noise in the solid-state medium \cite{beukers2024remote}.

Generating spin-spin entanglement based on interference of photons from distinct emitters requires long-term optical stability. Even for emitters that are excellent sources of indistinguishable photons in the short-term, any long-term spectral diffusion will lead to phase errors for the entangled state, resulting from the unknown frequency difference between emitters \cite{kambs2018limitations}.
While a known frequency difference can be compensated with quantum eraser techniques \cite{zhao2014entangling}, suppressing the effect of spectral diffusion requires post-selecting photons that arrive in a narrow time window \cite{bernien2013heralded,legero2003time,metz2008effect}: increasing fidelity by accepting smaller fractions of the dephasing time $T_{2}^*$ at the cost of reduced entanglement rates.

In this work, we propose a technique to correct phase errors in time-bin entanglement swapping protocols that arise from quasi-static spectral diffusion, an improvement upon a previously presented optical dynamical decoupling method \cite{ruskuc2023heralded}. The key idea is to rephase spectral diffusion by returning one of the emitters to the excited state after the entangled state is heralded, for a duration corresponding to the detection time of the heralding photons.
The final fidelity of the Bell state is independent of the magnitude of spectral diffusion, and only depends on the lifetime of the excited state used for shelving.
For systems where a long-lived shelving state is not available, the technique can be combined with post-selection for more favorable acceptance windows to increase fidelity, scaling with the optical lifetime $T_1$ rather than $T_{2}^*$.
For systems with long-lived shelving states, with lifetime much longer than the photon emission time, the phase error is corrected without reducing the entanglement rate.

Shelving states with correlated spectral diffusion can be obtained in systems with Kramers doublet excited states, such as T centers in Si \cite{higginbottom2022optical,johnston2024cavity}, V or Mo defects in SiC \cite{csore2020ab,bosma2018identification,wolfowicz2020vanadium,gilardoni2020spin} and Kramers' rare-earth ion defects (\emph{e.g.} Nd$^{3+}$, Er$^{3+}$ or Yb$^{3+}$) \cite{alqedra2023optical,dibos2018atomic,bottger2016optical}. 
Interfacing such systems with nanophotonic cavities can allow for selective Purcell enhancement of one of the excited state transitions, leaving a second excited state with a relatively longer lifetime for use as a shelving state.
A long-lived shelving state can also be achieved in 3-level systems with a single excited state using tunable cavities to dynamically change the emission rate of the excited state for emission or shelving \cite{yang2023controlling}.

In the following discussion, we describe the framework for our scheme, present single- and two-photon spin-spin entanglement protocols with rephasing and consider its dependence on the shelving state lifetime. Finally, we discuss several aspects of its use for \ercawo, which has recently shown high indistinguishability of its emitted photons \cite{ourari2023indistinguishable} and direct spin-photon entanglement in the telecom band \cite{uysal2024spin}. These demonstrations show that spectral diffusion observed over a long-term is largely quasi-static and can be corrected with the presented technique.
We expect that this technique can be an enabling tool for the generation of high-rate high-fidelity spin-spin entanglement based on solid-state emitters.

\begin{figure}[!t]
	\centering
    \includegraphics[width=86 mm]{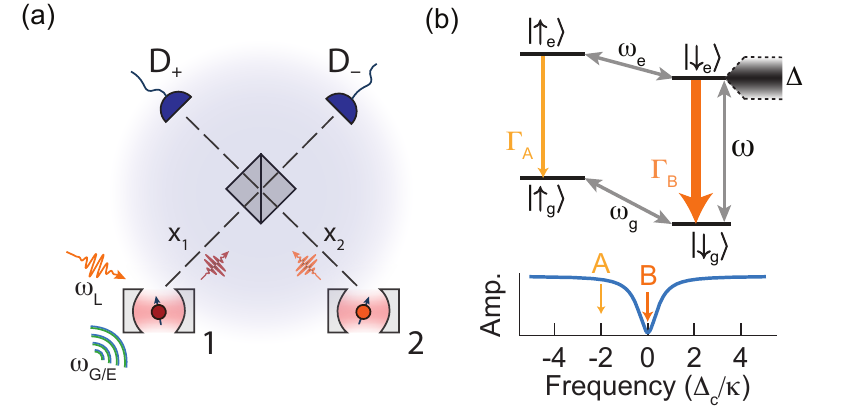}
    \caption{\label{fig:fig1}\textbf{Generating spin-spin entanglement.}
    \textbf{(a)} Two spins interfaced with an optical cavity, labeled 1 and 2, are excited by laser at frequency, $\omega_L$, and controlled using MW tones at $\omega_G$ and $\omega_E$, targeting the optical transition, ground and excited state splittings respectively. Photons from emitter-1(2) travel distance $x_1$($x_2$) before arriving at a 50:50 beam-splitter. Detection of a single photon at either detector $D_+$ or $D_-$ in given time-bins heralds entanglement.
    \textbf{(b)} 4-level system consists of ground and excited spin manifolds with frequency splittings $\omega_g$ and $\omega_e$ respectively. Spin-conserving transitions (labeled A,B) are highly cyclic, with spontaneous emission rate $\Gamma_B \gg \Gamma_A$ owing to tuning of the B transition on resonance with the cavity, and the detuning of the A transition away from it by a splitting $\Delta_{\textrm{split}}=|\omega_g-\omega_e|$, larger than the cavity linewidth $\kappa$. The B transition, addressed with the laser, has frequency $\omega$ and undergoes spectral diffusion with magnitude $\Delta$.}
\end{figure}

We consider two emitters in separate nodes, emitting photons that are overlapped on a central beamsplitter before detection  (Fig.~\ref{fig:fig1}a). A typical level structure for a Kramers' doublet system (\emph{i.e.}, \er) is shown in Fig.~\ref{fig:fig1}b, consisting of spin-1/2 ground and excited states with Zeeman splittings $\omega_{g,i}$ and $\omega_{e,i}$, and an optical transition frequency of $\omega_i$ for the indicated transition used for spin-photon entanglement (Fig.~\ref{fig:fig1}b). We take the spontaneous emission rate $\Gamma$ of the transition used for entanglement ($\Gamma_B$) to be  faster than that of the shelving state ($\Gamma_A$). As demonstrated in recent work \cite{ourari2023indistinguishable}, this can be achieved by selectively enhancing $\Gamma_B$ with an optical cavity (Fig.~\ref{fig:fig1}c). While the A and B transitions have different frequencies, we assume that their frequency fluctuations are correlated. This is a reasonable assumption in the context of Kramers' doublets in host materials with low magnetic noise, where the dominant noise is electric field noise and must act the same on the spin sublevels by Kramers' theorem \cite{kramers1930theorie}. The optical transitions are subject to slow spectral diffusion by an amount $\Delta_i = \omega_i - \bar{\omega}_i$. The lasers used to excite each ion are tuned to the average frequency, $\omega_{L,i} = \bar{\omega}_i$.

In the rotating frame for the ground and excited spin manifolds, we can describe the 4-level system in terms of the Hamiltonian ($\hbar=1$):
\begin{equation}
\label{eq:Hrot}
    H_i = \omega_{i} Z^o_i + \Delta_{g,i} Z^g_i + \Delta_{e,i} Z^e_i,
\end{equation}
where $Z^o_i = \mathrm{diag}(1,1,0,0)$ is the optical operator and $Z^g_i = \mathrm{diag}(0,0,1,0)$ and $Z^e_i = \mathrm{diag}(1,0,0,0)$ are ground and excited spin operators respectively. The ground and excited spin detunings are defined as $\Delta_{g,i}=\omega_{g,i}-\omega_{G,i}$ and $\Delta_{e,i}=\omega_{e,i}-\omega_{E,i}$ with respect to drive tones $\omega_{G,i}$ and $\omega_{E,i}$.

In addition to free evolution, spin-photon entanglement schemes also involve unitary operations on the spin and optical transitions. We define these operators in the spin-rotating frame as $R_i^j(\theta) = \exp(-i\theta X_i^j/2)$ for each emitter $i$, where $\theta$ is the angle of rotation and $X^j_i$ is the drive operator for each transition $j$ given as $X^{g/e}_i = \ket{\uparrow_{g/e}}\bra{\downarrow_{g/e}} +\ket{\downarrow_{g/e}}\bra{\uparrow_{g/e}}$ for the ground and excited spin transitions and $X^{o}_i = e^{-i\omega_{L,i} t} \ket{\downarrow_{e}}\bra{\downarrow_{g}} + e^{i\omega_{L,i} t}\ket{\downarrow_{g}}\bra{\downarrow_{e}}$ for the optical B transition. 

The final ingredient of spin-spin entanglement schemes is the spontaneous emission and subsequent detection of photons. Emitted photons can be described as occupying a spatio-temporal mode, whose shape is determined by the emitter properties \cite{legero2003time,kambs2018limitations}. A single photon in either mode ($i=1,2$) can be represented as $\zeta_i^*(z_i) a_i^{\dag}\ket{0_i}$, where $\ket{0_i}$ is the vacuum state, $a_i^{\dag}$ is the creation operator and $\zeta_i(z_i) = \sqrt{\Gamma_i}H(z_i)e^{-\Gamma_i z_i/2}e^{i\omega_i z_i}$ is the normalized field mode of the photons over the space-time coordinate $z_i = t-x_i/c$. $H(z_i)$ is the Heaviside function, which ensures that at $x_i=0$, the position of the emitter, the photon can only be observed after $t=0$, the time of excitation.
Correspondingly, we model the spontaneous emission of a photon as a quantum channel with Kraus operator: $C_{e,i} = \ket{\downarrow_g}\bra{\downarrow_e}\otimes \zeta_i^*(z) a_i^{\dag} + (I_e-\ket{\downarrow_e}\bra{\downarrow_e})\otimes I_p$, where $I_e$ is the identity operator for the emitter and $I_p$ is the identity operator for the photon channels. In effect, this channel maps the state $\ket{\downarrow_e}$ to $\ket{\downarrow_g}$, while generating a photon in mode $i$, with spatio-temporal shape and phase defined by $\zeta_i(z_i)$. To generate entanglement between spins, the emitted photons are overlapped on a beam-splitter with the unitary $U_{BS} = e^{i\pi/4(a_1a_2^{\dag}+a_1^{\dag}a_2)}$, removing the path information before detection \cite{beukers2024remote}. Equivalently, projectors can be defined in terms of symmetric and anti-symmetric combinations of the two modes, $\ket{\pm} = \ket{10}\pm\ket{01}$, where $\ket{10}$ ($\ket{01}$) is the joint state of the two photon modes with only a single photon in the first (second) mode. Then, a detection at time $\tau$ for detector, $D_{\pm}$ corresponds to projector $P_{\pm}(\tau) = \delta(t-\tau)\ket{\pm}\bra{\pm}$.

We assume $\Delta_i$ to be quasi-static, constant over the duration of each entanglement attempt but randomly distributed with linewidth $\sigma_f$ resulting in a dephasing time $T_2^* = 1/(\sqrt{2}\pi\sigma_f)$ for $\sigma_f \gg \Gamma$.
For simplicity, we set decay rates to be equal, $\Gamma_{1,2}=\Gamma=1/T_1$, and drop the subscript, $i$, on operators to denote the respective operators acting both emitters in parallel.

\begin{figure}[!t]
	\centering
    {\includegraphics[width=86 mm]{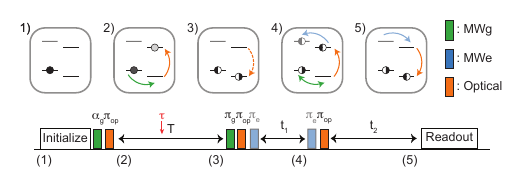}}
    \caption{\label{fig:fig2}\textbf{Single-photon scheme.}
    Top, depiction of the spin state for one of the emitters at the indicated points in the scheme. Bottom, pulse-sequence for the single-photon time-bin entanglement generation scheme (steps 1-3), and rephasing (steps 4-5). A single photon is detected at time $\tau$ with respect to the excitation.
    }
\end{figure}

Spin-spin entanglement can be implemented with single-photon \cite{cabrillo1999creation,bose1999proposal} and two-photon \cite{barrett2005efficient} detection protocols. We begin by considering the single-photon case (Fig.~\ref{fig:fig2}).
After initializing in $\ket{\uparrow_g\uparrow_g}$, a small angle MW rotation, $R^g(\alpha)$ with $\alpha \ll 1$, followed by an optical $\pi$-pulse, $R^o(\pi)$, yields the state: 
\begin{equation}
\begin{split}
    \ket{\psi}_{(S,2a)} =& R^o(\pi)R^g(\alpha) \ket{\psi}_{(S,1)} \\
    =& (1-\alpha)\ket{\uparrow_g\uparrow_g} \\
    &+\sqrt{\alpha(1-\alpha)}(\ket{\downarrow_e\uparrow_g}+\ket{\uparrow_g\downarrow_e})\\
    &+\alpha\ket{\downarrow_e\downarrow_e}.
\end{split}
\end{equation}

As the state decays, a photon is emitted by neither, one or both of the emitters depending on the spin states:
\begin{equation}
\begin{split}
    \ket{\psi}_{(S,2b)} =& C_e \ket{\psi}_{(S,2a)} \\
    =& (1-\alpha)\ket{\uparrow_g\uparrow_g}\otimes\ket{00} \\ 
     &+\sqrt{\alpha(1-\alpha)}\zeta_1^*(z_1)\ket{\downarrow_g\uparrow_g}\ket{10}\\
     &+\sqrt{\alpha(1-\alpha)}\zeta_2^*(z_2)\ket{\uparrow_g\downarrow_g}\ket{01})\\
     &+\alpha\zeta_1^*(z_1)\zeta_2^*(z_2)\ket{\downarrow_g\downarrow_g}\otimes\ket{11}.
\end{split}
\end{equation}
In the case that only one photon is emitted, we obtain the following Bell state after a time $T$:
\begin{equation}
\label{eq:standard_single_photon_bell}
\begin{split}
    \ket{\psi}_{(S,3)} =& \frac{1}{\sqrt{\alpha(1-\alpha)p(\tau)}}\int dt P_{\pm}(\tau)e^{-iHT}\ket{\psi}_{(S,2b)} \\
    \simeq& \frac{1}{\sqrt{2}}\left(e^{i(k_1x_1-\omega_1\tau)}e^{-i\Delta_{g,2}T}\ket{\downarrow_g\uparrow_g} \right. \\
    &\left. \pm e^{i(k_2x_2-\omega_2\tau)}e^{-i\Delta_{g,1}T}\ket{\uparrow_g\downarrow_g}\right)\\
    =& \frac{1}{\sqrt{2}}\left( e^{-i\phi_{\mathrm{S1}}}\ket{\downarrow_g\uparrow_g}\pm e^{-i\phi_{\mathrm{S2}}}\ket{\uparrow_g\downarrow_g}\right),
\end{split}
\end{equation}
where the sign of the relative phase depends on whether detector $D_+$ or $D_-$ received the photon at time $\tau$, the photon accumulates phase over the optical path length $x_i$ ($k_i = \omega_i/c$) and the spin accumulates phase over the waiting period $T$. The wavefunction is normalized by the probability density to detect a photon at time $\tau$, $p(\tau) = H(\tau-T_0)\Gamma e^{-\Gamma(\tau-T_0)}/2$, where $T_0 = (x_1+x_2)/(2c)$ is the mean time-of-flight. We have assumed that the difference in time-of flight, $\delta_T = \frac{x_2-x_1}{c}$, is small compared to the decay rate ($\delta_T \ll 1/\Gamma$).
 Due to photon loss in any physical system, the two-photon term $\ket{11}$ is also consistent with the heralding condition, so that the probability of having the Bell state in Eq.~\ref{eq:standard_single_photon_bell} given a photon click is $1-\alpha$ \cite{bose1999proposal}.
At this point, we consider the relative phase, $\phi_{\mathrm{S}} = \phi_{\mathrm{S2}}-\phi_{\mathrm{S1}}$, of the Bell state (Eq.~\ref{eq:standard_single_photon_bell}):
\begin{equation}
\label{eq:standard_single_photon_phase}
    \phi_{\mathrm{S}} = - \omega_0\delta_T - (\Delta_{g,2}-\Delta_{g,1})T  + (\omega_2-\omega_1)t_0,
\end{equation}
where $\omega_0=\frac{\omega_1+\omega_2}{2}$ is the mean frequency and $t_0 = \tau-T_0$ is the emission time from either emitter.

In the standard single-photon entanglement scheme, an unknown frequency difference $\omega_2-\omega_1$ leads to a phase error of the Bell state (Eq.~\ref{eq:standard_single_photon_phase}). Intuitively, this unknown phase shift corresponds to uncertainty about which emitter decayed. The phase error can be suppressed in a rate-fidelity trade-off that scales with $T_2^*$: accepting only the earliest events with $t_0$ ($t_0 \ll T_2^*$), at the cost of reducing the entanglement rate by a factor of $t_0 \Gamma$, which is significant if the dephasing time is short compared to the emission rate ($T_2^* \ll 1/\Gamma$).

The key idea of the rephasing protocol is to return the emitter to the excited state for a brief time \emph{after} the photon detection, once the emission time $t_0$ is known, to ensure each emitter accumulates the same optical phase, undoing the unwanted phase shift. The protocol consists of a $R^g(\pi)$ pulse followed by two $R^o(\pi)$ pulses separated by $t_1 = t_0 + \epsilon$ to shelve for a duration equal to the emission time, within error $\epsilon$ of misestimating the time of flight.
Provided the excited state does not decay during this protocol, this yields the state:
 \begin{equation}
\label{eq:post_single_photon_bell}
 \begin{split}
    |\psi\rangle_{(S,5)} =& R^o(\pi) e^{-iHt_1}R^o(\pi)R^g(\pi) \ket{\psi}_{(S,3)} \\
    =& \frac{1}{\sqrt{2}}\left( e^{-i\phi_{\mathrm{S1}}^{\prime}}\ket{\uparrow_g\downarrow_g}\pm e^{-i\phi_{\mathrm{S2}}^{\prime}}\ket{\downarrow_g\uparrow_g}\right),
\end{split}
\end{equation}
with corrected phase terms:
\begin{equation}
\label{eq:post_single_photon_phase_terms}
 \begin{split}
    \phi_{\mathrm{S1}}^{\prime} =&  \phi_{\mathrm{S1}} + (\Delta_{g,1}+\omega_2)t_1+\omega_{L,2}t_1,\\
    \phi_{\mathrm{S2}}^{\prime} =&  \phi_{\mathrm{S2}} + (\Delta_{g,2}+\omega_1)t_1+\omega_{L,1}t_1,
\end{split}
\end{equation}
where the laser phase ($\omega_{L,i}t_1$) between excitation and de-excitation is imprinted on the state. We again consider the relative phase, $\phi_{\mathrm{S}}^{\prime} = \phi_{\mathrm{S2}}^{\prime}-\phi_{\mathrm{S1}}^{\prime}$, after phase correction of the entangled state in Eq.~\ref{eq:post_single_photon_bell}:
\begin{equation}
\label{eq:post_single_photon_phase}
\begin{split}
    \phi_{\mathrm{S}}^{\prime} =& - \omega_0\delta_T - (\omega_{L,2}-\omega_{L,1})t_1 - (\omega_2-\omega_1)\epsilon\\
    &- (\Delta_{g,2}-\Delta_{g,1})(T-t_1) .
\end{split}
\end{equation}
Comparing Eq.~\ref{eq:post_single_photon_phase} with Eq.~\ref{eq:standard_single_photon_phase}, we have replaced the unknown phase term $(\omega_2-\omega_1)t_0$ with a fully known term $(\omega_{L,2}-\omega_{L,1})t_1$ that depends on the optical control frequencies rather than the emitter frequencies. The residual error term, which still depends on the ion frequency difference, is insignificant provided that the average time of flight is known to sufficient precision ($\omega_2-\omega_1 \ll 1/\epsilon$). The term $\omega_0\delta_T$ indicates that path length fluctuations also lead to a phase error, as is standard for single-photon schemes \cite{beukers2024remote}. Provided that the difference in path lengths is small in relation to spectral diffusion ($\delta_T \ll 1/\Delta_i$), this term can be corrected by stabilizing or monitoring fluctuations in $\delta_T$. For example, spectral diffusion near 200 kHz \cite{ourari2023indistinguishable} would require a path-length difference much less than 100~m. Finally, we note that waiting an additional time $t_2 = T-t_1$ after the shelving step also corrects for any unknown variations in $\Delta_{g,i}$ indicated by the last term in Eq.~\ref{eq:post_single_photon_phase}.

A spontaneous decay during the shelving step will destroy the entangled state. However, such a decay is unlikely if the emission time $t_0$ is small in comparison to the lifetime, $T_{1}$, changing the time-scale of the rate-fidelity trade-off from $T_2^*$ to $T_1$.

This challenge can be overcome if a long-lived shelving state with correlated spectral diffusion is available. In this case, the post-selection window can be extended further to include all emission events without a cost to fidelity in the limit of much longer shelving lifetimes. 
In this case, we modify the sequence by adding two $R^e(\pi)$ pulses (Fig.~\ref{fig:fig2}), to instead accumulate phase in the shelving state for the correction (Appendix~\ref{sec:corr_shelve}). 

A phase error in the two-photon entanglement scheme \cite{barrett2005efficient} can be similarly corrected. In the two-photon scheme a photon is emitted by both emitters, and the unknown phase shift instead depends on the difference in arrival times of the two detected photons. To correct the phase error, we can similarly shelve in the excited state for a duration equal to the arrival time difference (Appendix~\ref{sec:two_photon}).


\begin{figure*}[!t]
	\centering
    {\includegraphics[width=172 mm]{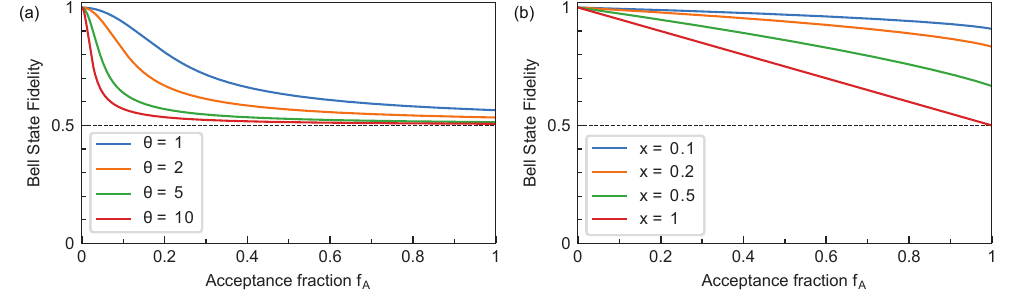}}
    \caption{\label{fig:fig3}\textbf{Entanglement fidelity.}
    \textbf{(a)} Post-selection fidelity, $F_{\textrm{PS}}(\theta,f_A)$ and \textbf{(b)} rephasing fidelity $F_{\textrm{RP}}(x,f_A)$ for several values of the spectral diffusion ratio, $\theta$ and the shelving lifetime ratio, $x$, is plotted as a function of $f_A$.
    }
\end{figure*}

Next, we evaluate how the fidelity of the resultant state depends on the lifetime of the shelving state used. While the best performance is obtained with long-lived shelving states, the rephasing scheme can be favorable even in the limit where the shelving state does not provide a longer lifetime, in comparison to the loss of fidelity imposed by spectral diffusion.

To investigate the lifetime dependence, we allow for spontaneous emission during the shelving step. Then, instead of the pure Bell state, $\ket{\psi_{\mathrm{Bell}}}$, in either the single- or two-photon schemes (Eqs.~\ref{eq:post_single_photon_bell_alt},\ref{eq:psi_T8a}), we get the mixed state:
\begin{equation}
    \label{eq:mixed_postphasing}
    \rho = (1-e^{-\Gamma_A t_1}) \ket{\uparrow_g\uparrow_g}\bra{\uparrow_g\uparrow_g}
    +  e^{-\Gamma_A t_1}\ket{\psi_{\mathrm{Bell}}}\bra{\psi_{\mathrm{Bell}}},
\end{equation}
where $e^{-\Gamma_A t_1}$ is the probability of not decaying from the shelving state within required shelving time $t_1$. We can then calculate the Bell state fidelity as $F_{\textrm{RP}}(t_1)=\bra{\psi_{\mathrm{Bell}}}\rho\ket{\psi_{\mathrm{Bell}}} = e^{-\Gamma_A t_1}$, which is simply the probability of not decaying during the shelving step. Furthermore, the shelving time $t_1$ is distributed according to probability density $p(t_1) = \Gamma_Be^{-\Gamma_B t_1}$, based on the probability density of single photon clicks for the single-photon scheme or the time-difference of photon clicks for the two-photon scheme. When using the rephasing scheme, we can always limit our operation to a given cut-off time, $t_c$, such that we only accept events where $t_1 < t_c$. Equivalently, the cut-off time corresponds to an acceptance fraction of all events, $f_A = 1-e^{-\Gamma_Bt_c}$, based on the probability density, $p(t_1)$. Then, we can calculate the fidelity of the entangled state as a function of $f_A$:
\begin{equation}
    \label{eq:analytic_fidelity}
    F_{\textrm{RP}}(x,f_A) = \int_0^{t_c} dt_1 p(t_1)F_{\textrm{RP}}(t_1) = \frac{1-(1-f_A)^{1+x}}{f_A(1+x)},
\end{equation}
where $x = \Gamma_A/\Gamma_B$ is the ratio of the shelving state decay rate to the excited state used to generate photons. 
If all events are accepted ($f_A=1$), the expression simplifies to $F_{\textrm{RP}}(x) = 1/(1+x)$, showing an inverse relation with the lifetime ratio. Alternatively, for an equal lifetime-shelving state ($x=1$), we find $F_{\textrm{RP}}(f_A) = 1-f_A/2$, linearly decreasing with the acceptance fraction.

We can compare the rephasing fidelity with post-selection of smaller time windows without rephasing. For two emitters with variance $\sigma_f^2$ of spectral diffusion, the fidelity of the Bell state given $t_1$ can be calculated as $F_{\textrm{PS}}(t_1) = (1+e^{-\beta^2t_1^2/2})/2$, where $\beta = \sqrt{2}\times 2\pi\sigma_f$, by integrating the Bell state overlap
over the frequency distributions \cite{kambs2018limitations}. We can similarly calculate the fidelity for a given acceptance fraction $f_A$:
\begin{equation}
\label{eq:fidelity_post_selection}
\begin{split}
    F_{\textrm{PS}}(\theta,f_A) &= \int_0^{t_c} dt_1 p(t_1)F_{\textrm{PS}}(t_1)\\
    &= \frac{1}{2}\left( 1+\int_0^{f_A} df e^{-(2\pi\theta\log(1-f))^2}\right),
\end{split}
\end{equation}
where $\theta = \sigma_f/\Gamma$ is the ratio of spectral diffusion to the decay rate and the integral can be calculated numerically. In Fig.~\ref{fig:fig3}, we compare rephasing, $F_{\textrm{RP}}(x,f_A)$, and post-selection, $F_{\textrm{PS}}(\theta,f_A)$, fidelities for several values of $x$ and $\theta$. When the spectral diffusion is 10 times the decay rate, the acceptance fraction needs to be reduced significantly to reach high entanglement fidelities using post-selection (Fig.~\ref{fig:fig3}a). In contrast, any amount of spectral diffusion can be corrected with rephasing, where the fidelity only depends on the shelving lifetime (Fig.~\ref{fig:fig3}b).
We note that faster frequency fluctuations, not considered here, would reduce the fidelity of both approaches and could be compensated by decreasing the acceptance fraction further for either approach.


We now discuss several aspects of implementing our scheme. Like with the standard schemes, the experimental delay $T$ for the rephasing scheme must be at least twice the communication delay $T_0$, so that information on detector click patterns and photon detection times can be relayed back to each node for local operations. 
For accurate correction of the Bell state, the detection time resolution, $t_{\textrm{res}}$, must be high enough to resolve the ion frequency difference and spectral diffusion, satisfying $t_{\textrm{res}} \ll 1/|\omega_2-\omega_1|,1/\Delta_i$. For \ercawo, spectral diffusion is only on the order of 100~kHz, while the inhomogenous distribution is less than 1~GHz \cite{ourari2023indistinguishable}, so that $t_{\textrm{res}}\sim10-100$~ps, easily achievable with superconducting nanowire single-photon detectors (SNSPD) \cite{allmaras2019intrinsic}, is sufficient. Finally, the lifetime ratio between the optical transitions can be calculated, for a cavity on resonance with the optical $B$ transition, as $x = (1 + (2\Delta_{\textrm{split}}/\kappa)^2)^{-1}$, where $\Delta_{\textrm{split}} = |\omega_g-\omega_e|$ is the splitting between the $A$ and $B$ transitions and $\kappa$ is the cavity linewidth \cite{dibos2018atomic}. We can obtain a lifetime ratio of $x=0.1$ for a cavity linewidth of $\kappa\sim 1$~GHz and $\Delta_{\textrm{split}}\sim1.5$~GHz, which is achievable at a magnetic field strength $|B|\sim1000$~G given the \er ion ground and excited state g-tensors \cite{ourari2023indistinguishable}. While we expect that the long-lived shelving state will be correlated in energy owing to the Kramers' doublet structure of the excited state in \ercawo, we note that it still needs to be verified experimentally.

Here, we have presented a protocol to correct phase errors due to frequency fluctuations in the quasi-static regime for the single- and two-photon spin-spin entanglement schemes. The requirement for a long-lived shelving state can be satisfied by rare earth ions or color centers in Si and SiC with a Kramers doublet level structure and selective Purcell enhancement of optical transitions using nanophotonic cavities. For 3-level systems where a correlated shelving state is not available, the scheme can be combined with post-selection to achieve more favorable acceptance fractions depending on the magnitude of long-term spectral diffusion. The rephasing scheme could also be combined with tunable nanohotonic devices \cite{yang2023controlling} to dynamically change the Purcell enhancement for shelving or emission, removing the need for an additional long-lifetime shelving state. We expect that the scheme can be an enabling tool for generating high-rate entanglement for solid-state emitters, which need to contend with spectral diffusion in the solid-state medium. 
Together with recent demonstrations of indistinguishable photons and spin-photon entanglement in \ercawo, we expect that our scheme will allow demonstrations of long-distance high fidelity spin-spin entanglement at the telecom band in the near future.

We acknowledge Andrei Faraon and Andrei Ruskuc for discussions about related rephasing approaches, as well as helpful conversations with Sebastian Horvath, Haitong Xu, Lukasz Dusanowski, Salim Ourari, Adam Turflinger.

This work was primarily supported by a DOE Early Career award (DE-SC0020120). Additional support was provided by the U.S. Department of Energy, Office of Science, National Quantum Information Science Research Centers, Co-design Center for Quantum Advantage (C2QA) under contract number DE-SC0012704, and AFOSR (FA9550-18-1-0334).

\emph{Note:} While finalizing this manuscript, we became aware of related work using a related method \cite{ruskuc2024scalable}.

\appendix

\section{Using a shelving state}
\label{sec:corr_shelve}

Here, we briefly discuss the rephasing sequence making use of a shelving state with correlated spectral diffusion for the phase correction, indicated by the addition of the excited state $\pi$ pulses in Fig.~\ref{fig:fig2}.

After the end of the single-photon protocol, the Bell state $\ket{\psi}_{(S,3)}$ (Eq.~\ref{eq:standard_single_photon_bell}) is generated. To make use of the shelving state for rephasing, we apply a $R^o(\pi)$ pulse followed by a $R^e(\pi)$ pulse and shelve in the second excited state for a duration $t_1$:

 \begin{equation}
\label{eq:alt_S4}
\begin{split}
    \Tilde{|\psi\rangle}_{(S,4)} =& e^{-iHt_1}R^e(\pi)R^o(\pi)R^g(\pi) \ket{\psi}_{(S,3)} \\
    =& \frac{1}{\sqrt{2}}\left( e^{-i\phi_{\mathrm{S1}}^{\prime}-i\omega_{L,2}T}e^{-i(\Delta_{g,1}+\Delta_{e,2}+\omega_2)t_1 } \ket{\uparrow_g\uparrow_e} \right. \\
    &\left. \pm e^{-i\phi_{\mathrm{S2}}^{\prime}-i\omega_{L,1}T}e^{-i(\Delta_{g,2}+\Delta_{e,1}+\omega_1)t_1 }\ket{\uparrow_e\uparrow_g} \right) .
\end{split}
\end{equation}

At this point, we have accumulated the required excited state phase to correct for the phase error caused by spectral diffusion. As discussed in the main-text, this assumption is supported by the Kramers' theorem, which states that energy of a Kramers doublet must be degenerate in the presence of electric fields. However, we have also acquired phase under magnetic field fluctuations, given by the ground and excited state detunings, $\Delta_{g,i}$ and $\Delta_{e,i}$. While the ground and excited state spin levels have different sensitivity to magnetic fields, the phase each level accumulates under a magnetic field should still be correlated by the ratio $r$ of their magnetic moments, given as $r = \omega_E/\omega_G$. We use this correlation to cancel these terms subsequently in the ground state, by applying a $R^e(\pi)$ pulse followed by a $R^o(\pi)$ pulse to return to the ground state and waiting here for a modified duration $t_2 = T - t_1(1-r)$:
 \begin{equation}
\label{eq:alt_S5}
\begin{split}
    \Tilde{|\psi\rangle}_{(S,5)} =& e^{-iHt_2}R^o(\pi)R^e(\pi) \Tilde{|\psi\rangle}_{(S,4)} \\
    =& \frac{1}{\sqrt{2}}\left( e^{-i\Tilde{\phi}_{\mathrm{S1}}^{\prime}}\ket{\uparrow_g\downarrow_g} 
    \pm e^{-i\Tilde{\phi}_{\mathrm{S2}}^{\prime}}\ket{\downarrow_g\uparrow_g} \right) ,
\end{split}
\end{equation}
with the corrected phase terms:
\begin{equation}
\label{eq:post_single_photon_bell_alt}
 \begin{split}
    \Tilde{\phi}_{\mathrm{S1}}^{\prime} =&  \phi_{\mathrm{S1}} + (\Delta_{g,1}+\Delta_{e,2}+\omega_2)t_1+\omega_{L,2}t_1 + \Delta_{g,1}t_2,\\
    \Tilde{\phi}_{\mathrm{S2}}^{\prime} =&  \phi_{\mathrm{S2}} + (\Delta_{g,2}+\Delta_{e,1}+\omega_1)t_1+\omega_{L,1}t_1 + \Delta_{g,2}t_2,
\end{split}
\end{equation}

Under the assumption that the detunings caused by magnetic field fluctuations are also correlated by $\Delta_{e,i}/\Delta_{g,i} = r$, we obtain the relative phase $\Tilde{\phi}_{\mathrm{S}}^{\prime} = \Tilde{\phi}_{\mathrm{S2}}^{\prime}-\Tilde{\phi}_{\mathrm{S1}}^{\prime}$:
\begin{equation}
\label{eq:post_single_photon_phase_alt}
    \Tilde{\phi}_{\mathrm{S}}^{\prime} = - \omega_0\delta_T - (\omega_{L,2}-\omega_{L,1})t_1 - (\omega_2-\omega_1)\epsilon .
\end{equation}
After the correction, the quasi-static frequency fluctuations are still fully canceled. We note that the assumed correlation within the Kramers doublet excited states still need to be verified experimentally.

\section{Two-photon scheme}
\label{sec:two_photon}

\begin{figure*}[!ht]
	\centering
    {\includegraphics[width=172 mm]{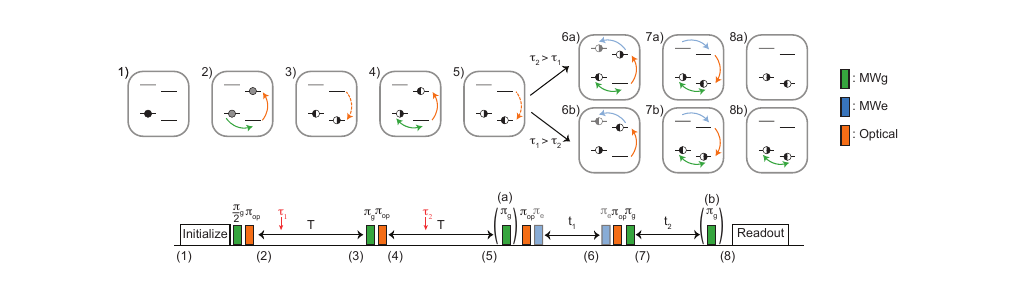}}
    \caption{\label{fig:fig4}\textbf{Two-photon scheme.}
    Top, depiction of the spin state for one of the emitters at the indicated points in the scheme. Bottom, pulse-sequence for the two-photon time-bin entanglement generation scheme (steps 1-5), and rephasing (steps 6-8). A single photon is detected at times $\tau_1$ and $\tau_2$ with respect to excitations in each window.
    }
\end{figure*}

We now discuss the two-photon time-bin entanglement scheme \cite{barrett2005efficient} (Fig.~\ref{fig:fig4}). We again start with the initialized state,
\begin{equation}
    \ket{\psi}_{(T,1)} = \ket{\uparrow_g\uparrow_g}.
\end{equation}
We then apply a $R^g(\pi/2)$ pulse to prepare a spin superposition, followed by a $R^o(\pi)$ pulse to excite,
\begin{equation}
\label{eq:psi_T2}
\begin{split}
    \ket{\psi}_{(T,2)} =& R^o(\pi)R^g(\pi/2)\ket{\psi}_{(T,1)} \\
    =& \frac{1}{2}\left( \ket{\uparrow_g\uparrow_g}+\ket{\downarrow_e\uparrow_g}+\ket{\uparrow_g\downarrow_e}+\ket{\downarrow_e\downarrow_e}\right) .
\end{split}
\end{equation}
In the same manner as the single-photon scheme, the emission of a single photon and its detection after some time, $\tau_1$, with respect to the first excitation pulse at either detector projects the state into:
\begin{equation}
\begin{split}
    \ket{\psi}_{(T,3)} =& \frac{1}{\sqrt{p(\tau)/4}}\int dt P_{\pm}(\tau_1)e^{-iHT}C_e\ket{\psi}_{(T,2)}\\
     =& \frac{1}{\sqrt{2}}\left( e^{i(k_1x_1-\omega_1\tau)}e^{-i\Delta_{g,2}T}\ket{\downarrow_g\uparrow_g}\right. \\
    &\left. \pm e^{i(k_2x_2-\omega_2\tau)}e^{-i\Delta_{g,1}T}\ket{\uparrow_g\downarrow_g}\right) ,
\end{split}
\end{equation}
However in contrast to the single-photon scheme, which makes use of a weak superposition, the detection of a single photon here is equally likely to correspond to the term $\ket{\downarrow_e\downarrow_e}$ in Eq.~\ref{eq:psi_T2}, which emits two photons. In the two-photon scheme, this possibility is eliminated by applying a $R^g(\pi)$ pulse to flip the spin state and performing a second optical excitation, $R^o(\pi)$,
\begin{equation}
\begin{split}
    \ket{\psi}_{(T,4)} =& R^o(\pi)R^g(\pi)\ket{\psi}_{(T,4)}\\
    =&\frac{1}{\sqrt{2}}\left( e^{i(k_1x_1-\omega_1\tau)}e^{-i\Delta_{g,2}T}e^{-i\omega_{L,2} T}\ket{\uparrow_g\downarrow_e}\right. \\
    &\left. \pm e^{i(k_2x_2-\omega_2\tau)}e^{-i\Delta_{g,1}T}e^{-i\omega_{L,1} T}\ket{\downarrow_e\uparrow_g}\right) .
\end{split}
\end{equation}
The detection of a second photon at time $\tau_2$ with respect to the second excitation heralds entanglement and yields,
\begin{equation}
\label{eq:psi_T5}
\begin{split}
    |\psi\rangle_{(T,5)} =& \frac{1}{\sqrt{p(\tau)/2}}\int dt P_{\pm}(\tau_2)e^{-iHT}C_e\ket{\psi}_{(T,4)}\\
    =& \frac{1}{\sqrt{2}}\left(e ^{-i\phi_{\mathrm{T1}}}\ket{\uparrow_g\downarrow_g}\pm e^{-i\phi_{\mathrm{T2}}}\ket{\downarrow_g\uparrow_g}\right) ,\\
\end{split}
\end{equation}
where the phase terms $\phi_{T1}$ and $\phi_{T2}$ are defined as:
\begin{equation}
\label{eq:phi_T5}
\begin{split}
    \phi_{T1} =& \phi_{T0} + \omega_1\tau_1+\omega_2\tau_2+\omega_{L,2}T,\\
    \phi_{T2} =& \phi_{T0} + \omega_2\tau_1+\omega_1\tau_2+\omega_{L,1}T.
\end{split}
\end{equation}
Here, the global phase is given as $\phi_{T0} = -(k_1x_1+k_2x_2)+(\Delta_{g,1}+\Delta_{g,2})T$. The sign of the second term is ($+$) if both photons were detected on the same detector and ($-$) if detected on different detectors.
We again consider the relative phase of the Bell state, $\phi_{\mathrm{T}} = \phi_{\mathrm{T2}}-\phi_{\mathrm{T1}}$, (Eq.~\ref{eq:psi_T5}) after the standard two photon time-bin entanglement protocol:
\begin{equation}
\label{eq:standard_two_photon_phase}
    \phi_{\mathrm{T}} = -(\omega_2-\omega_1)(\tau_2-\tau_1)-(\omega_{L,2}-\omega_{L,1})T.
\end{equation}

For the two-photon scheme, the unknown frequency difference $\omega_2-\omega_1$ similarly leads to a phase error of the Bell state (Eq.~\ref{eq:standard_two_photon_phase}). In this case, a photon is emitted by both emitters, either first or second for either term in the Bell state (Eq.~\ref{eq:psi_T5}), and the relative phase corresponds to the assignment of detection times $\tau_i$ to the emitters.

To correct the optical phase, we shelve in the excited for a duration given by the arrival time difference of photons, similarly ensuring that each emitter accumulates the same optical phase. Here, we first make a decision on whether $\tau_1$ or $\tau_2$ is greater. If $\tau_2 > \tau_1$, this implies that the second excitation accumulated phase for a longer time and we follow the (a) branch in Fig.~\ref{fig:fig4}. To correct for this difference, we apply a $R^g(\pi)$ pulse followed by two $R^o(\pi)$ pulses separated by $t_1 = |\tau_2 - \tau_1|$:
\begin{equation}
\label{eq:psi_T8a}
\begin{split}
    |\psi\rangle_{(T,8)} =& e^{-iHt_2}R^g(\pi)R^o(\pi)e^{-iHt_1}R^o(\pi)R^g(\pi) |\psi\rangle_{(T,5)}\\
    =& \frac{1}{\sqrt{2}}\left( e^{-i\phi_{\mathrm{T1}}^{\prime}}\ket{\uparrow_g\downarrow_g}
    \pm e^{-i\phi_{\mathrm{T2}}^{\prime}}\ket{\downarrow_g\uparrow_g}\right) ,
\end{split}
\end{equation}
where the corrected phase terms are:
\begin{equation}
\label{eq:phi_T7a}
\begin{split}
    \phi_{\mathrm{T1}}^{\prime} =&  \phi_{\mathrm{T1}}+(\omega_1+\Delta_{g,2})t_1 + \Delta_{g,1}t_2 - \omega_{L,1}t_1\\
    \phi_{\mathrm{T2}}^{\prime} =&  \phi_{\mathrm{T2}} +(\omega_2+\Delta_{g,}1)t_1 + \Delta_{g,2}t_2 - \omega_{L,2}t_1,
\end{split}
\end{equation}
and we have waited for an additional time $t_2 = t_1$ to also refocus the spin-phase accumulated during the correction step.
In the (b) branch, where $\tau_1 > \tau_2$, we do not apply the $R^g(\pi)$ pulse in step (6b) to accumulate relative phase at the opposite rate, but instead apply it in step (8b) to obtain the same state. In either case, we arrive at the final state with relative phase $\phi_{\mathrm{T}}^{\prime} = \phi_{\mathrm{T2}}^{\prime} - \phi_{\mathrm{T1}}^{\prime}$:
\begin{equation}
\label{eq:post_two_photon_phase}
    \phi_{\mathrm{T}}^{\prime} = -(\omega_{L,2}-\omega_{L,1})(\tau_2-\tau_1)-(\omega_{L,2}-\omega_{L,1})T.
\end{equation}
After rephasing, the phase is entirely known.
The remaining dependency on the optical control frequencies and photon arrival times can be accounted by subsequent ground state operations. 

A long-lived shelving state can be used by inserting $R^e(\pi)$ pulses in between the two $R^o(\pi)$ pulses.  Similar to the single-photon scheme, the second waiting time $t_2$, needs to be modified such that $t_2 = t_1|1-r|$ in order to account for the excited state splitting. If we work with a level structure where $\omega_g > \omega_e$, this is the only required change and the sequence is exactly as presented in Fig.~\ref{fig:fig4}. If instead we have $\omega_e > \omega_g$, the ground $\pi$-pulse in step 7 also needs to be omitted to account for the different sign of the spin phase.

\bibliography{post-phasing.bib}

\end{document}